# Gender Performance in Physics Practicals

**Maria Vivien Visaya and Tasneem Limbada**

Department of Pure and Applied Mathematics, University of Johannesburg

mvvisaya@uj.ac.za

**Abstract**. An analysis of longitudinal student Physics practicals data over a period of three semesters is conducted. In particular, we study the association of the gender of students and their Physics practicals marks in mechanics, thermodynamics, optics, electricity, and overall marks (OVM). Together with gender, all variables are binarized, i.e. distinction mark (≥75) =1 and zero otherwise. To visualize performance of students, the qualitative method of plotting a two-dimensional orbit is used to represent binary multivariate longitudinal data of each student. Analysis of orbits reveals information of patterns in the data. This study gives a good indication of which Physics practical each gender group perform better in. Our visual analysis indicate that male students tend not to get distinctions in both mechanics and OVM, but that females tend to get distinctions in OVM. We have also observed that students' 0/1 marks is most stable (i.e. least changing) in optics and OVM, and most frequently changing in electricity. A comparison to the GEE statistical model indicates that visual results present initial insights to help and complement statistical analysis.

## 1. Introduction

We investigate the association of gender of students to four Physics practicals marks and their overall mark. Differences in male and female performance in Physics have been discussed in literature. In [1], it is argued that there is no correlation between male and female student's performance in Physics practicals, but that male students performed better in Physics practical than female. Results in [2] show that male students performed better than females using both conventional laboratory apparatus and the micro science kits. On the other hand [3, 4] show that gender has no significant effect on student's practical skills in Physics. Findings in [5] reveal that gender is not a determinant of students' academic performance in Physics.

The variables to be considered in our analysis are student gender, practical marks in mechanics, thermodynamics, optics, and electricity, and overall practical average mark (OVM, the average of the four practicals). The following six binary questions (and associated variables with 0/1 coding) are used in our analysis. We note that $Q_2$ to $Q_5$ are asked per semester.

$Q_0$ (G) : Is the student a male (M) or female (F)? (F=0, M=1)

$Q_1$ (OVM): Did the student achieve a distinction IN OVM? (YES=1)

$Q_2$ (Mech): Did the student achieve a distinction in mechanics? (YES=1)

$Q_3$ (Therm): Did the student achieve a distinction in thermodynamics? (YES=1)

$Q_4$ (Opt): Did the student achieve a distinction in optics? (YES=1)

$Q_5$ (Elec): Did the student achieve a distinction in electricity? (YES=1)

## 2. The Method of Orbits

Longitudinal data is interesting as it tracks the same type of information on the same subject at multiple points in time [6]. Understanding the changes that have occurred in a longitudinal study over time, particularly those with huge amount of quantitative information, are easier to interpret and understand in their visual format [7, 8].

For visual analysis of multivariate longitudinal data, very few are available when data is binary. The method of orbits [9] is a simple way of visualizing such data but useful in handling data sets with large number of variables and subjects, in either long or short period of time. For consistent interpretation of visuals, coding of answers is either 1 if favorable, or 0 otherwise. The detailed description of the orbit method is given in [9] and applied in [10] but we briefly discuss the construction of orbit below.

### 2.1 Orbit Algorithm

*Step 1:* Determine the initial state $\boldsymbol{p_0^k} = (\boldsymbol{x_0^k}, \boldsymbol{y_0^k})$ of subject *k*. For each subject *k* in the population, let $\boldsymbol{f_{ij}^k}$ be the frequency of change in answer of question/variable $i_j$ over the observation period. Suppose

$$0 < f_{i0}^k < f_{i1}^k < \dots < f_{ij}^k < f_{i(n-1)}^k,$$

the initial question order of $k$ is $y_0^k = i_0 i_1 \dots i_{n-1}$. If $f_{ij}^k = f_{ij+1}^k$ and $f_{ij}^{pop} < f_{ij+1}^{pop}$ at the population level, then question order is chosen as $i_j i_{j+1}$. If $f_{ij}^{pop} = f_{ij+1}^{pop}$, then question order is according to the trivial order 012... $(n-1)$. The initial answer state of *k* is

$$\boldsymbol{x_0^k} = \boldsymbol{x_0 x \dots x_{n-1}}$$

where $x_j$ is the corresponding answer to variable $i_j$ in $y_0^k$. The initial state of *k* is $p_0^k = (x_0^k, y_0^k)$.

*Step 2:* The next states $\boldsymbol{p_t^k} = (\boldsymbol{x_t^k}, \boldsymbol{y_t^k})_{\boldsymbol{t>0}}$ of a subject *k* is as follows.
Identify the question/s that change answer values at t=1. If there are none, then the next state $\boldsymbol{p_1^k} = \boldsymbol{p_0^k}$. If the answer of variable $\boldsymbol{i_j}$ changes at t=1, then swap to the rightmost position both $\boldsymbol{i_j}$ (in $\boldsymbol{y_0^k}$) and $\boldsymbol{x_j}$ (in $\boldsymbol{x_0^k}$) to the rightmost position and change $\boldsymbol{x_j}$ to its new value $\boldsymbol{x_j^*}$. Suppose that more than one answer changes at t=1. Then sequentially swap changing questions and corresponding new answers to the right, starting with the rightmost changing question. The new order of answers and variables comprise $(\boldsymbol{x_1^k}, \boldsymbol{y_1^k}) = \boldsymbol{p_1^k}$.
*Step 3*: Iterate Step 2 until end of time period T.
*Step 4*: The orbit of *k* is the sequence of states $O(k) = \boldsymbol{p_0^k}, \boldsymbol{p_1^k}, \dots, \boldsymbol{p_T^k}$, where $\boldsymbol{p_t^k}, \boldsymbol{p_{t+1}^k}$ are connected by an edge.

Note that the orbit algorithm is applied to each subject in the study. The method simply rearranges data, together with variables, at each time step so no information is lost nor added. For analysis in *n* variables, orbits are plotted in a 2-dimensional space $\boldsymbol{S_n} = \boldsymbol{X_n} \times \boldsymbol{Y_n}$ where $\boldsymbol{X_n}$ is the set of all possible concatenated answers and $\boldsymbol{Y_n}$ is the set of all concatenated order of variables. Table 1 gives an example of a binary longitudinal data in 3 variables together with the states $x_t$ and $y_t$ of the orbit. We can trace the raw data from columns $x_t$ and $y_t$ For instance, at t=0, we know that variables1, 2 and 0 have values 1, 1, and 0 respectively. The asterisk indicates that variable zero changes answer at t=1 from 0 to 1.

**Table 1.** Binary coding of data for variables $Q_0$ , $Q_1$, and $Q_2$. Values in the states ($x_t$,$y_t$) at each time t is such that order of variables in $y_t$ give corresponding answer in $x_t$. Values with an asterisk change value in the next time step while red/bold numbers in $y_t$ correspond to variable i (=$Q_i$) that changed value in the next time step.

| t | $Q_0$ | $Q_1$ | $Q_2$ | $x_t$ | $y_t$ |
|---|---|---|---|---|---|
| 0 | 0* | 1 | 1 | 110* | 12**0** |
| 1 | 1* | 1 | 1* | 11*1* | 1**20** |
| 2 | 0 | 1 | 0 | 100 | 102 |

## 3. Student Data

Our longitudinal data is composed of 74 students (25 males and 49 females) taking analytical chemistry and food technology at the Doornfontein Campus, University of Johannesburg. Information was gathered over a period of 3 semesters for each student. Only students with data for all four practicals in mechanics, thermodynamics, optics, and electricity for all three semesters are considered. All students fall under the extended program course in Physics practical. Students have been split into groups with a specific demonstrator and demonstrators are changed every semester. The marker uses a special Excel program designed to mark experiments.

## 4. Results

Figure 1 illustrates orbits of all 74 students in the 2-d state space $S_6$. State transitions (edges) that go to the left are colored red while edge transitions that goes to the right are colored green. As a consequence of the swapping operation in constructing orbits, a red edge is considered going to an "unfitter" state (i.e. decrease in the number of 1's) while green is considered going to a "fitter" state (i.e. increase in the number of 1's ). Edge transitions with the same x-value (i.e. same number of 1's and 0's as the previous semester) are colored blue.

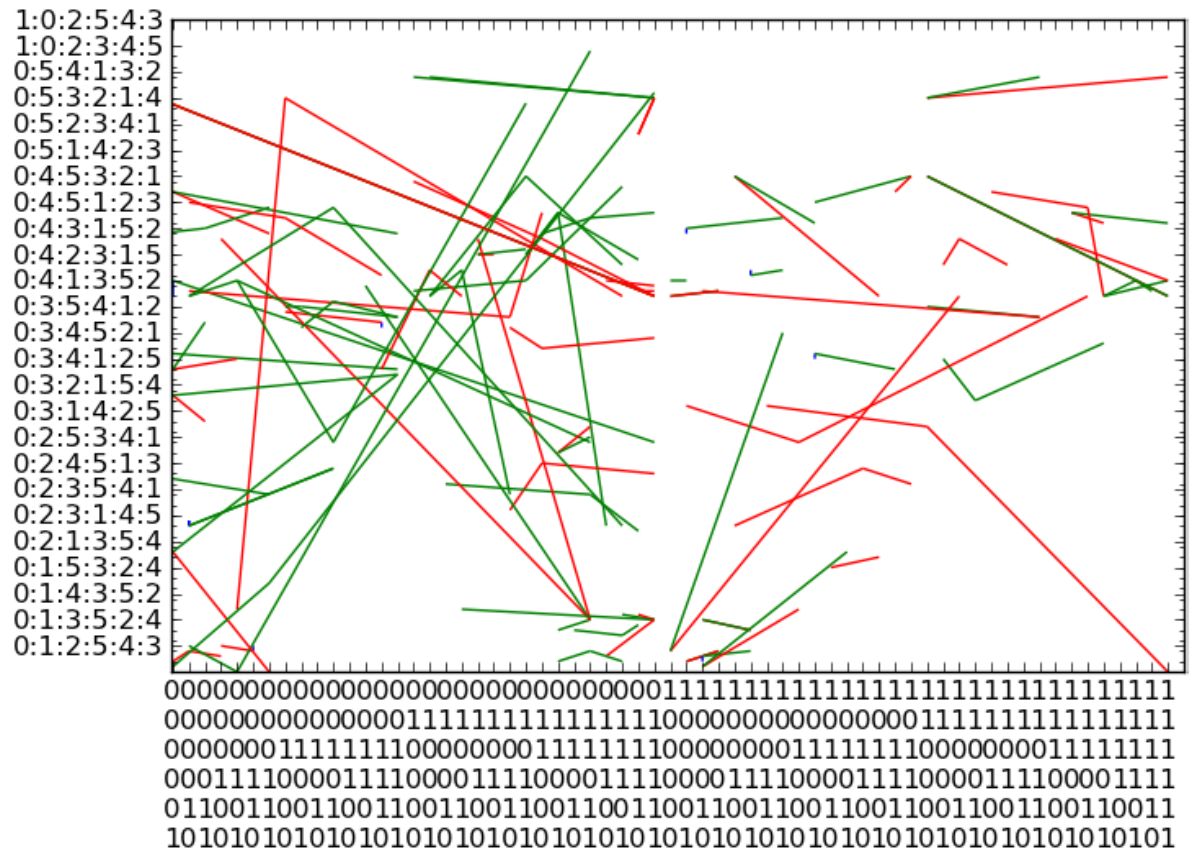

**Figure 1.** Orbits of all 74 students (25 males and 49 females) automatically cluster along y=0****, with left-most variable 0 associated to $Q_0$ (gender). The cluster on the left are female orbits, on the right are male.

From the orbits in Figure 1, we can clearly see students who have performed good or bad (distinction-wise) during the 3 semester period. Extreme transitions are illustrated (i.e. a large increase/decrease in the number of 0's or 1's). For instance, in the female group, we see an orbit composed of two green edges- starting at (000010,012543) where only variable 4 (optics)=1 but ends at (011011, 102345) where all are distinction except for variable 3(therm). In the male group we see an orbit consisting of two red edges that starts at (111111,012345) where it is distinction in all variables, and ends at (100110,031425) where only variables 4 (optics) and 2(mechanics) are distinction.

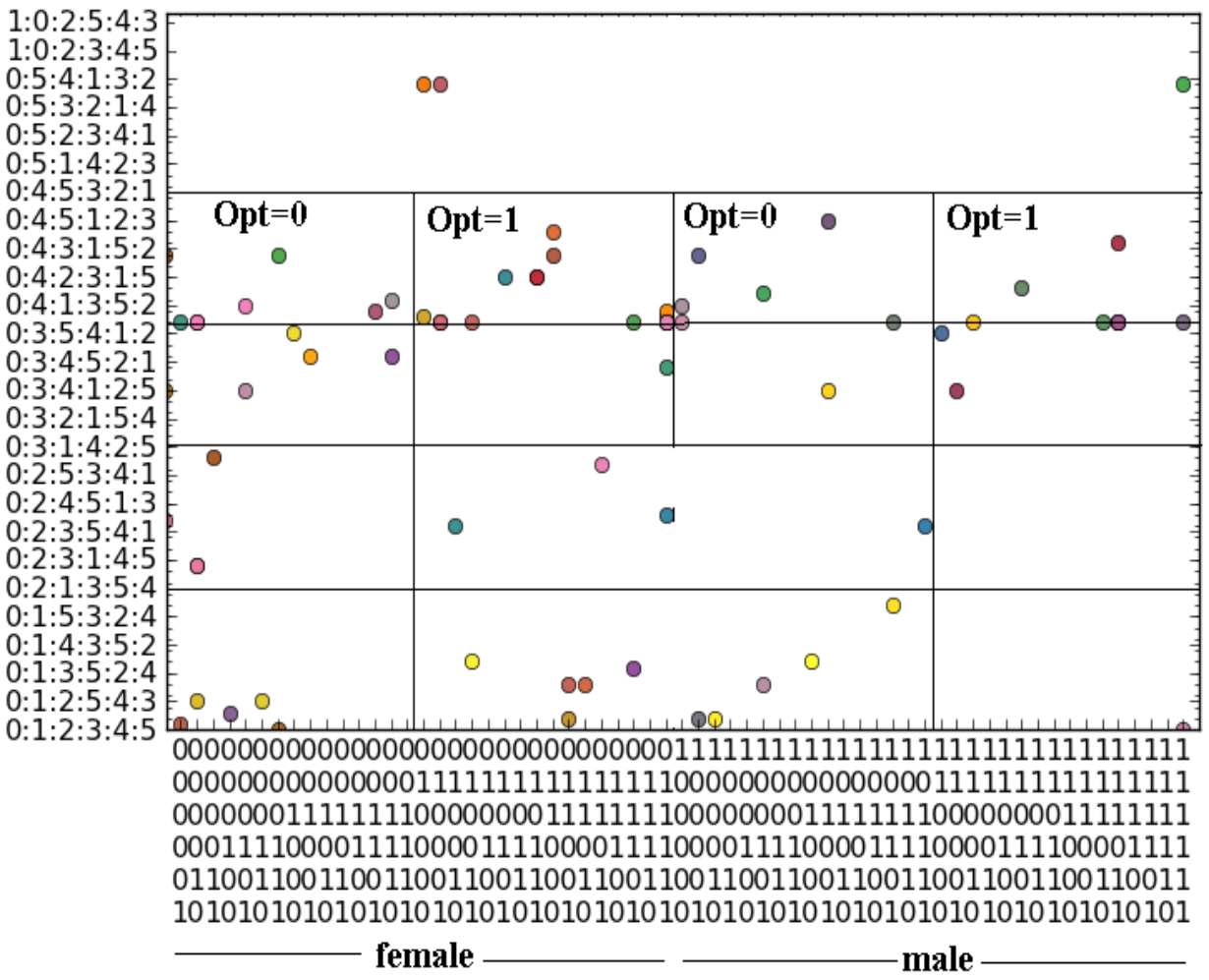

**Figure 2.** Initial condition of orbits. Most orbits start at y=04**** or y=01**** which means that the 0/1 mark of most students is least changing in $Q_4$(Optics) and $Q_1$(OVM).

Figure 2 illustrates the initial condition of orbits. We see that most orbits start at y=04**** or y=01****. This means that the 0/1 mark of most students is least changing in $\boldsymbol{Q_4}$(Optics) and $\boldsymbol{Q_1}$(OVM). This is verified by the change frequency $f_i$ over three semesters for each variable $i$ for all 74 students:

$$f_0(\text{G}) : 0,\ f_4(\text{opt}) : 64,\ f_1(\text{OVM}) : 68,\ f_2(\text{mech}) : 72,\ f_3(\text{therm}) : 73,\ f_5(\text{elec}) : 89.$$

To give us an idea about the density of orbit visits, we illustrate in Figure 3 the heat map of male and female orbits. In Figure 3 we drop the gender variable so variable 0 is now OVM, variable 1 is now mechanic, and so on. It is clear that females tend to get more distinction in OVM as seen in the density of visits in OVM=1. On the other hand, males tend *not* to get distinction in both Mechanics and their overall mark, as seen in the lack of visit in MECH=1 and OVM=1.

For a comparison of our visual results to statistical results , we use a generalized estimating equation model (GEE), the most applicable statistical method to use for our type of longitudinal (repeated) binary data [11]. We see from Table 2 that females are more likely to get a distinction in OVM, MECH, and OPT (odds ratio <1) and male more likely to get distinction in thermodynamics and electricity (odds ratio>1). However only OVM is significant, and is agreement with our orbit observation (Figure 3).

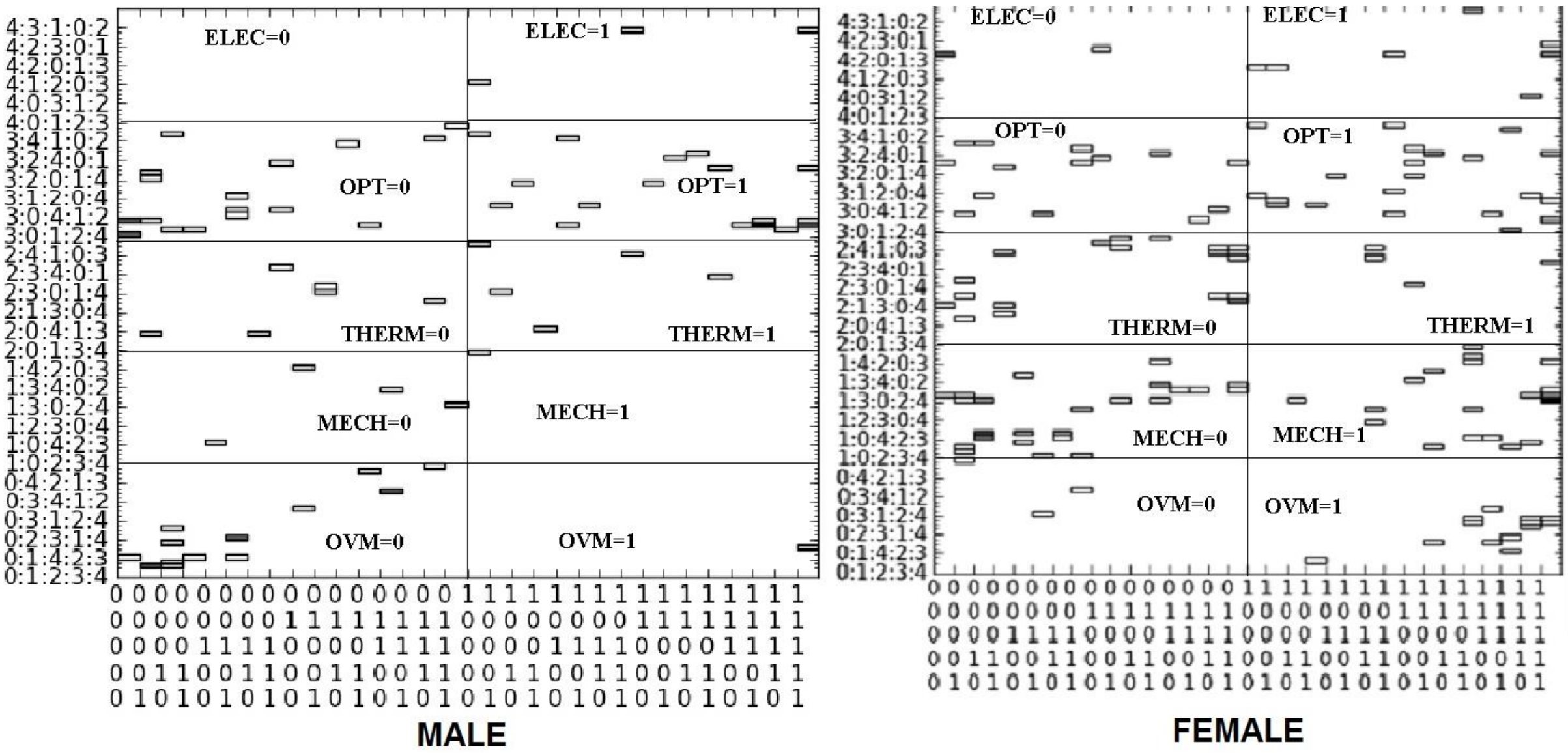

**Figure 3.** Heat map of male and female orbits denoting the density of visits to each state in the state space. The darker the state, the more visit by orbits.

**Table 2. Information for the GEE model. A significant value of 0.198 (and odds ratio of 0.646 <1) implies that males are less likely to get a distinction in OVM compared to females.**

| | | | | | |
|---|---|---|---|---|---|
| Dependent Variable : gender=1 | | | Subject Effect : ID | | |
| Probability Distribution : Binomial Link | | | Within-Subject Effect : Semester | | |
| Function : Logit | | | Working Correlation Matrix Structure : AR(1) | | |
| Parameters | Significance | Odds Ratio | Parameters | Significanc e | Odds Ratio |
| (Intercept) | .198 | .646 | [THERM=1] | .448 | 1.262 |
| [OVM=1] | .013 | .367 | [OPTICS=1] | .466 | .798 |
| [MECH=1] | .629 | .852 | [ELEC=1.00] | .187 | 1.575 |

## 5. Concluding Remarks

We analyzed student data in six variables by using a graphical display of binary multivariate longitudinal data. Orbits have been useful for the three-semester short duration data as we can immediately see interesting transitions and the population and individual level (extreme ones, for instance). Visual inspection of orbits can help in hypothesis construction in statistical analysis. A comparison to the GEE statistical model tells us that visual results present initial insights to help and complement statistical analysis. We note that our specific data was particularly taken from students taking food technology and analytical chemistry so results may differ for students who from a different course (e.g. in the Engineering). In addition, since there are different laboratory demonstrators per semester, the authors also recommend analysis per semester. A quick modification of the GEE model in Table 1 is given in Table 2 where the “semester” variable is included as a predictor.

**Table 2.** Model information for the GEE model including semester, where semester 1 is the reference category (and only the significant OVM is shown). Among the 4 practicals and OVM, only OVM is (again) significant at 0.006, with odds ratio (OR) = 0.348 which implies that males are less likely to get a distinction in OVM compared to females. Semester 3 is significant with OR=1.366 which means there is significant difference in performance of male over female in semester 3 compared to semester 1.

| Parameter | Sig. | Odds Ratio |
|---|---|---|
| [OVM=1.00] | .006 | .348 |
| [SEMESTER=3] | .085 | 1.366 |
| [SEMESTER=2] | .978 | .996 |